\documentclass[preprint,10pt]{elsarticle}
\usepackage{setspace} 
\usepackage{graphicx}
 \usepackage{amssymb}
\usepackage[ruled,vlined]{algorithm2e}

\newtheorem{defi}{Definition}
\newtheorem{lemma}{Lemma}
\newtheorem{theorem}{{\bf Theorem}}

\newtheorem{prob1}{Problem}

\begin{document}
\begin{frontmatter}
\title{Pattern Formation for Asynchronous Robots without Agreement in Chirality}
\author{Sruti Gan Chaudhuri$^1$,  Swapnil Ghike$^2$, Shrainik Jain$^3$ and Krishnendu Mukhopadhyaya$^4$}
\address{$^1$ Information Technology, Jadavpur University, Kolkata - 700032, India. Email: srutigan@gmail.com}
\address{$^2$ LinkedIn, Mountain View, CA - 94043, USA. email: swapnil.ghike@gmail.com}
\address{$^3$ University of Washington, USA, email: shrainik@gmail.com}
\address{$^4$ ACM Unit, Indian Statistical Institute, Kolkata - 700108, India. Email: krishnendu@isical.ac.in}

\begin{abstract}
This paper presents a deterministic algorithm for forming a given asymmetric pattern in finite time by a set of {\it autonomous}, {\it homogeneous}, {\it oblivious} mobile robots under the CORDA model. The robots are represented as points on the 2D plane. 
There is no explicit
communication between the robots. The robots coordinate among themselves by observing
the positions of the other robots on the plane. 
Initially all the robots are assumed to be stationary. The robots have local coordinate systems defined by Sense of Direction (SoD), orientation or chirality and scale. 
Initially the robots are in asymmetric configuration. We show that these robots can form any given asymmetric pattern in finite time.

\end{abstract}
\end{frontmatter}

\section{Introduction}
Executing a collaborative task by a set of small, autonomous, mobile robots (also known as {\it robot swarm} \cite{Peleg2005}) has been a popular topic for the research for last few decades. The robots are assumed to be anonymous, homogeneous, oblivious and asynchronous.
The robots together perform a complex job, e.g., moving a big body \cite{Noreils1993}, cleaning a big surface \cite{Jung1998} etc. 
The individual unit or robot in a system of swarm robots is less expensive than a big robot. Increasing or decreasing the number of robots in this system involves very simple hardware or software modifications and thus provides good scalability. Moreover, having similar capability, if some robots fail, others can manage to execute the
work. This feature makes the system to be more resilient to malfunction. In hostile environments, these robots are easily deployable to perform various complex tasks cooperatively. 

An important task for a set of mobile robots is \emph{pattern formation} \cite{Sugihara1990}. A set of robots is required to arrange themselves on a 2D plane to form a pattern, in finite time. The robots are considered as points. Initially all robots are at distinct positions on the 2D plane. If the robots can form any pattern, they can agree on their subsequent roles in future coordinated action. In this paper we propose a deterministic distributed algorithm for a given asymmetric pattern formation by a set of asynchronous mobile robots. 

\subsection{Earlier Works}
An extensive volume of research \cite{Agmon2004, Balch1998, Beni1991, Efrima2009, Flocchini2000, Flocchini2008, 
Gordon2008, Katayama2007, Klasing2008, Parker2000, Prencipe2007} has been reported in the context of multiple autonomous
mobile robots exhibiting cooperative behavior for the swarm robots. Many of them are based on geometric pattern formation. 
There exists several algorithms for forming specific patterns such as circle \cite{Defago2008}, straight line \cite{Prencipe2006} etc. Arbitrary pattern formation is a general case of pattern formation. The pattern is usually given as an input described either in the form of a set of points expressed by their mutual distances and angles, or as geometric figures like polygon, circle, straight line etc. 
A complete characterization of the class of formable patterns \cite{Suzuki1993,Suzuki1996,Suzuki1999} has been reported for the robots under Sync and SSync model and also when the robots have an unbounded amount of memory. 
Flocchini et al. \cite{Flocchini1999,PFlocchini2001,Flocchini2008} investigated arbitrary pattern formation problem for asynchronous and oblivious robots for different cases where robots may or may not have common SoD and/or chirality. With common SoD and common chirality any pattern is formable for any number of robots \cite{PFlocchini2001}. With common SoD and without common chirality any pattern may be formed with odd number of robots \cite{Flocchini1999}. If the number of robots is even, they can form {\it symmetric patterns} \cite{Flocchini1999}.  
Without common SoD, the robots can not form an arbitrary pattern even if they agree on chirality \cite{PFlocchini2001}.
They also showed that without common SoD, a set of initial configurations, known as {\it symmetric configurations}, exists, for which it is impossible to form an arbitrary pattern \cite{Flocchini2008}.
Ghike and Mukhopadhyaya \cite{Ghike2010} presented a deterministic algorithm for a given pattern formation without SoD and Chirality. Their solution finds collision free paths for the robots. In order to achieve this, some robots are selected for moving at a time. However, their solution assumed that no tie occurs when a robot is selected for movement. They also assumed that the robots on the Smallest Enclosing Circle (SEC) of the robots are less than or equal to the number of robots inside the SEC. 
In this paper we propose algorithms for arbitrary pattern formation for points robots which eliminates these limitations in \cite{Ghike2010}.

\section{Overview of the problem}
This paper presents an algorithm for formation of arbitrary asymmetric patterns by point robots under the CORDA model. The features of the robots are described as follows:
\begin{itemize}
\item Robots are autonomous, anonymous and homogeneous in the sense that
they are not uniquely identifiable, neither with a unique
identification number nor with some external distinctive mark (e.g., color,
flag, etc.). 
\item They are represented as points on the 2D plane. Robots have no common coordinate system. Each robot uses its own
local coordinate system defined by its origin, Sense of Direction (SoD), orientation or chirality and scale or unit distance. A robot has no knowledge about the coordinate system of any other robot.
\item Robots can not communicate explicitly. Each robot has a
camera/sensor which can take picture or sense over 360 degrees. The robots
communicate only by
means of observing other robots using the camera/sensor. A robot can compute
the coordinates (w.r.t. its own coordinate system) of other robots by observing
through the camera/sensor.  
\item Robots have infinite visibility range , i.e., a robot can see all other robots.
\item Robots execute the cycle ({\it Wait-Look-Compute-Move}) asynchronously. A robot does nothing in {\it Wait} state. In {\it Look} state, it gets the positional information of other robots by observing its surroundings. In {\it Compute} state it computes a destination point to move to. Finally in {\it Move} state, it moves to the computed destination along a straight line.  
\item  Under the CORDA model \cite {Prencipe2001}, the movement of the robots is not instantaneous. While in motion,
a robot may be observed by other robots. A robot may also stop before reaching its destination.
\item Robots are oblivious. They do not retain the information from the previous
cycles. 
\item Initially all robots are stationary. 
\end{itemize}

Let $\bar{r} = \{r_1, r_2, \ldots, r_n\}$ be a set of robots. Initially, $\bar{r}$ is assumed to be in asymmetric configuration \cite{Gan2013}\footnote{In asymmetric configuration there exists no straight line which divides the set of robots into two halves such that one half is the mirror image of the other.}. Our algorithm finds collision free paths (non-intersecting paths) for all the robots such that finally the robots form the given pattern.

A pattern $\cal P$ is defined by a set of $n$ points represented by their coordinate values with respect to an arbitrary coordinate system. i.e., ${\cal P} = \{(x_i,y_i) : 1 \le i \le n\}$.
Formally the problem is stated as follows: 
\begin{prob1}
\label{prob_p}
We are given a set of robots $\bar{r}$ which are in  asymmetric configuration and an asymmetric pattern $\cal P$. 
The robots in $\bar{r}$ have to move themselves to form $\cal P$ in finite time. 
\end{prob1}

\section{Solution approach}

This section gradually builds an algorithm to form the input pattern $\cal P$ by the robots in $\bar{r}$. The pattern formation algorithm has multiple sub-algorithms. We describe them one by one. 

\subsection{Agreement in coordinate system}
An important issue in the problem is representation of the given pattern. The pattern, given, is defined by an arbitrary coordinate system and the robots interpret it in their own coordinate system. An agreement in coordinate system with respect to the given pattern is found such that the representation of the pattern is same for all the robots. {\bf AgreementPattern()} does this job. 
First the $SEC$ of $\cal P$ is constructed. Let $c_{\cal P}$ be the center of the $SEC$. $c_{\cal P}$ becomes the common origin, for the pattern. 
We fix an ordering $Ord(\cal P)$ for the point in $\cal P$. 
A point, $p_l \in \cal P$, is selected so that it is the first point on the $SEC$ of $\cal P$, in $Ord(\cal P)$. 

\begin{lemma}
\label{lem-plfound}
 It is possible to elect a leader $p_l \in \cal P$ such that $p_l$ is on the $SEC$ of $\cal P$. 
\end{lemma}
\emph{Proof:}
$\cal P$ is asymmetric and hence orderable. If we fix an ordering and from that ordering choose the first point that is on the $SEC$ of $\cal P$, we shall have a leader lying on the $SEC$ of $\cal P$. 
\qed

$|c_{\cal P}p_{l}|$ is the common unit distance, $\overrightarrow{c_{\cal P}p_{l}}$ is the common positive $X$ axis for the pattern. Let $p'_l \in {\cal P}$ be a point  on $SEC$, which is presented next to $p_l$ in $Ord(\cal P)$. 

\begin{lemma}
\label{lem-p'lfound}
It is possible to select a point $p'_l \in \cal P$, different from $p_l$, such that $p'_l$ is on the $SEC$ of $\cal P$. 
\end{lemma}
\emph{Proof:}
 $\cal P$ is asymmetric and hence orderable. If we fix an ordering and from that ordering choose the first non leader point, $p'_l$, such that it does not lie on $\overline{c_{\cal P}p_l}$. 
 Since, the SEC contains at least two points existence of such a point is guaranteed.
\qed

The side of $X$ axis where $p'_l$ lies is considered as the side of the common positive $Y$ axis for the pattern. Note that the algorithm {\bf AgreementPattern()} also normalizes the pattern by representing the radius of the $SEC$ as the unit distance. 

\begin{algorithm}[H]
\KwIn{ $\cal P$: the set of pattern points.}
\KwOut{ A common  center, unit distance and axes of $\cal P$.}
Compute $SEC$ of $\cal P$;
$c_{\cal P} \leftarrow$ center of $SEC$\;
$Ord(\cal P) \leftarrow$ ordering of $\cal P$\;
$p_l \leftarrow$ the first point from $SEC$ in $Ord(\cal P)$\;
+ve $X$ axis $\leftarrow$  $\overline{c_{\cal P}p_l}$\;
$p'_l \leftarrow$ a point next to $p_l$ in $Ord(\cal P)$ such that it does not lie on $\overline{c_{\cal P}p_l}$\;
+ve $Y$ axis $\leftarrow$ the perpendicular of $\overline{c_{\cal P}p_l}$, drawn at $c_{\cal P}$, at that side of +ve $X$ axis, where $p'_l$ lies\;  
Return $c_{\cal P}$ as center, $|c_{\cal P}p_l|$ as unit distance, +ve $X$ and +ve $Y$\;
\caption{AgreementPattern()} 
\label{AgreementPattern}
\end{algorithm}

\paragraph{\bf Correctness of AgreementPattern():} The correctness of the algorithm follows from lemma \ref{lem-agreement}.
\begin{lemma}
\label{lem-agreement}
The origin, unit distance and axes are uniquely defined by {\bf AgreementPattern()}.
\end{lemma}
\emph{Proof:}
$SEC$ of $\cal P$ is unique. Hence, $c_{\cal P}$, i.e., the origin is unique.
Lemmas \ref{lem-plfound} and \ref{lem-p'lfound} ensure that, $p_l$ and $p'_l$ are unique. Hence, orientation of the axes (+ve $Y$ axis w.r.t. the +ve $X$) are unique.   
\qed

\vspace{30pt}
Using algorithm {\bf AgreementCoordinateSystem()}, the robots plot $\cal P$ in their local coordinate systems and fix common origin, axes and scale. The pattern formation algorithm is designed in such a way that, the agreement in coordinate system remains unchanged till the formation of $\cal P$ by $\bar{r}$ is complete. 
Algorithm {\bf AgreementCoordinateSystem()} first computes the $SEC$ of $\bar{r}$. Let $c_{\bar{r}}$ be the center of the $SEC$. 
An ordering $Ord(\bar{r})$ is fixed for $\bar{r}$.
The first robot, $r_l$, in $Ord(\bar{r})$, which is lying on $SEC$ of $\bar{r}$, is selected as leader.

\begin{lemma}
\label{lem-rlfound}
 It is possible to select an $r_l \in \bar{r}$ from $SEC$ of $\bar{r}$.
\end{lemma}
\emph{Proof:}
 Since, $\bar{r}$ is in  asymmetric configuration, it is orderable. If we fix an ordering, $r_l$ may be chosen as the first robot in that ordering which lies on the $SEC$ of $\bar{r}$. 
\qed

\vspace{20pt}
$\cal P$ is plotted so that $c_{\bar{r}} = c_{\cal P}$ and $r_l = p_l$. $c_{\bar{r}}$ becomes the common origin, denoted by $O$. $\overline{c_{\cal P}p_l}$ or $\overline{c_{\bar{r}}r_l}$ becomes the positive $X$ axis. The positive $Y$ axis for $\cal P$ is the common positive $Y$ axis for the robots. $|c_{\bar{r}}r_l| = |c_{\cal P}p_l|$ is the common unit distance, denoted by $u$. The other points in $\cal P$ are plotted accordingly. ${\cal P}*$ is the set of the coordinate values of the pattern points computed in the defined coordinate system.  

\begin{algorithm}[H]
\KwIn{$\bar{r}$: the set of robots, $\cal P$: the set of pattern points.}
\KwOut{$O$: Common origin, $XY$ axes, $u$: unit distance for the robots in $\bar{r}$, and ${\cal P}*$: set of coordinates of the pattern points.}
Compute $SEC$ of $\bar{r}$\; 
$c_{\bar{r}} \leftarrow$ Center of $SEC$ of $\bar{r}$\;
$Ord(\bar{r}) \leftarrow$ an ordering of $\bar{r}$\;
$r_l \leftarrow$ first robot in $Ord(\bar{r})$\;
AgreementPattern()\;
Plot $\cal P$ such that:\\ 
$\  \ \ $ $O \leftarrow c_{\bar{r}} \leftarrow c_{\cal P}$\; 
$\  \ \ $ $r_l \leftarrow p_l$\;
$\  \ \ $ $u \leftarrow |Or_l|$\;
$\  \ \ $ +ve $X$ axis for $\bar{r} \leftarrow \overline{c_{\bar{r}}r_l}$\;
$\  \ \ $ +ve $Y$ axis for $\bar{r} \leftarrow$ +ve $Y$ axis for $\bar{r}$\;
Compute all pattern points in $\cal P$\;
${\cal P}* \leftarrow$ set of coordinates of the pattern points\;
Return $O$, +ve $X$ and $Y$ for $\bar{r}$, $u$, ${\cal P}*$\;
\caption{AgreementCoordinateSystem()}
\label{Agreementcoordinatesystem}
\end{algorithm}

\paragraph{\bf Correctness of AgreementCoordinateSystem():}
Algorithm {\bf AgreementCoordinateSystem()} ensures that all robots agree on the orientation and scale of pattern to be formed. The correctness of the algorithm follows from lemma \ref{lem-agree-points}.
Since the coordinate system has been uniquely defined for $\bar{r}$, we can state the following lemma.
\begin{lemma}
\label{lem-agree-points}
{\bf AgreementCoordinateSystem()} computes the coordinate values of all pattern points in $\cal P$ uniquely and the computation is invariant of the position of the robots in $\bar{r}$. 
\end{lemma}

\subsection{Pattern formation}
Note that, the algorithms described so far do not require any robot to move. Once a robot fixes the coordinate axes and the pattern points, it is ready to move. However, the movements are designed in such a way that a robot that starts late will have the same coordinate system. This is ensured by maintaining the $SEC$ of the robots and the leader in the initial configuration remains the leader. The formation of pattern $\cal P$ by the robots in $\bar{r}$ is carried out through the following steps: 

\begin{itemize}
 
\item {\bf Step 1.} If $O \notin {\cal P}*$ and $\exists$ a robot $r_0$ at $O$, then $r_0$ moves by distance $\epsilon < d$ (where $d$ is the distance of a nearest robot not at $O$, from $O$) in the direction of positive $X$ axis.
\item {\bf Step 2.} If $O \in {\cal P}*$ and $\nexists$ a robot at $O$, a robot nearest to $O$ moves to $O$. Tie, if any, is broken using $Ord(\bar{r})$.
\item {\bf Step 3.} Let $p_{1}$ be the pattern point nearest to $O$ (if there are many, we choose the first one in $Ord(\cal P)$). $r_1$ is the robot nearest to $O$ (if there are many, we choose the first one in $Ord(\bar{r})$). Let $d=max(|Or_{1}|,|Op_1|)$. 
The robots which lie inside or on $Cir(O, d)$, move radially to a distance $d + \epsilon$ from $O$.
\item {\bf Step 4.} $r_{1}$ moves to $p_{1}$.
\item {\bf Step 5.} If $\exists$ free pattern points in ${\cal P}*$ on the $SEC$, they are filled as follows:
\begin{itemize}
\item If $\exists$ free robots inside $SEC$, then these robots move to fill the points in ${\cal P}*$ on the boundary of the $SEC$.
\item Else, robots on the $SEC$ move to occupy free ${\cal P}*$ on $SEC$ (without changing $SEC$ itself).
\end{itemize}
\item {\bf Step 6.} The rest of the robots which are not in position in ${\cal P}*$  move to occupy the free points in ${\cal P}*$.
\end{itemize}

Algorithm {\bf MoveRadiallyOut($r_i$)} executes the $3^{rd}$ step of the above list of operations. This algorithm also assures that there will be no collision between robots during  movements. The algorithm assumes that the robots agree in coordinate system (using {\bf AgreementCoordinateSystem()}).

\begin{algorithm}[H]
\KwIn{$\bar{r}$: the set of robots, $\cal P$: the set of pattern points.}
\KwOut{ A new configuration for the robots such that the executing robot lies at a distance $d+\epsilon$ from $O$.}
Find $r_{1}$ and $ p_{1}$\;
$d \leftarrow max(|Or_{1}|,|Op_{1}|)$\; 
$\bar{r'} \leftarrow $ $\{r_k \in \bar{r}$: s.t. $r_k$ lies inside $Cir(O,d)$\}\;
Compute $Ord(\bar{r})$\;
$r_i \in \bar{r'} \leftarrow$ robot nearest to the boundary of $Cir(O, d+\epsilon)$ (in case of tie use $Ord(\bar{r})$)\;
{
$h \leftarrow$ intersection of $\overline{Or_i}$ and  $Cir(O, d+\epsilon)$\; 
\eIf{$h$ is not occupied by any robot}
{$r_i$ moves to $h$\;}
{$g \leftarrow$ a point at a side of $h$ which is not occupied by other robot and $\angle{gr_ih}\le 90^o$\;
$r_i$ moves to $g$\;}
}
\caption{MoveRadiallyOut($r_i$)}
\label{Moveradiallyout}
\end{algorithm}

\paragraph{\bf Correctness of MoveRadiallyOut($r_i$):}
The correctness of the algorithm is established by lemma \ref{lem-radi-out}.

\begin{lemma}
\label{lem-radi-out}
{\bf MoveRadiallyOut($r_i$)} ensures collision free movement of the robots.
\end{lemma}

\emph{Proof:}
{\bf MoveRadiallyOut($r_i$)} checks if the robot executing the algorithm is inside the circle of radius $d$. If the robot is inside the circle, then it moves $\epsilon$ distance radially outward. To do so, first the robot nearest to the boundary of  $Cir(O,d+\epsilon)$ is identified.
There may be more than one such robots. In order to avoid possible collisions between robots the algorithm selects one robot for moving. The robot which is nearest to the boundary of  $Cir(O,d+\epsilon)$ and comes first in $Ord(\bar{r})$ is selected for moving. After selecting the robot for movement, the algorithm finds the destination for movement. The destination is selected in such a way that, it is not already occupied by other robots. Robots being points, a point, not occupied by other robot, will always exist on the boundary of $Cir(O, d+\epsilon)$. Moreover, during the movement towards destination, the robot remains nearest to the boundary of $Cir(O, d+\epsilon)$. Thus collisions with other robots are avoided.  
\qed

{\bf MoveRadiallyOut($r_i$)} also makes sure that all robots lie in the annular region between the SEC and $Cir(O,d)$. 
This also ensures that the subsequent movement $r_{1}$ to $p_{1}$ in step 4 is collision free.

\vspace{30pt}
Now we describe algorithm {\bf MoveToDestination($r_i$)} which finds destinations for each robot in $\bar{r}$, such that the paths of the robots to their respective destinations are collision free. The algorithm assumes that the robots agree in coordinate system (using {\bf AgreementCoordinateSystem()}).

\begin{algorithm}[H]
\KwIn{$\bar{r}$: the set of robots, $\cal P$: the set of pattern points.}
\KwOut{Destination for each robot in ${\cal P}*$.}
$FreeP \leftarrow$ \{$p_j \in {\cal P}*$: $p_j$ is not occupied by any robot\}\;
$Free\bar{r} \leftarrow$ \{$r_j \in \bar{r}$: $r_j \notin {\cal P}*$\}\;
${\cal S} \leftarrow \{(|Or_j|$, $\angle{r_jOp_k}):  r_j \in Free\bar{r},$ $p_k \in FreeP\}$\;
Let $(|Or_0|, \angle{r_0Op_0})$ be the lexicographic minimum of $\cal S$\;
$d' \leftarrow |Op_1|$\;
\eIf{$r_i \neq r_0$}
{
$Destination \leftarrow null$\;
}
{
\eIf{$\overline{r_0p_0}$ does not intersect $Cir(O, d'+\epsilon)$}{
\eIf{no other robots lie on $\overline{r_0p_0}$}{
$Destination \leftarrow p_0$\;
}
{
$SafeRegion \leftarrow$ the part of the region inside the $SEC$ but outside $Cir(O, d'+\epsilon)$ that lies between $\overrightarrow{Or_0}$ and $\overrightarrow{Op_0}$ s.t. $\angle{r_0Op_0} \leq 180^o$\;
$d_0 \leftarrow$ a point in $SafeRegion$, s.t. there exists no robots on $\overline{d_0p_0}$\;
$Destination \leftarrow d_0$\;
}
}
{
Find ${\cal T}^1_{r_0}$, ${\cal T}^2_{r_0}$, the tangents from $r_0$ to $Cir(O, d'+\epsilon)$ at points $t^1_{r_0}$ and $t^2_{r_0}$ respectively \;
Find ${\cal T}^1_{p_0}$, ${\cal T}^2_{p_0}$, the tangents from $p_0$ to $Cir(O, d'+\epsilon)$ at points $t^1_{p_0}$ and $t^2_{p_0}$ respectively\;
 ${\cal K} \leftarrow$ the set of points of intersection of ${\cal T}^i_{r_0}$,${\cal T}^j_{p_0}$ for $1 \le i,j,\le 2$\;
Find $d_1 \in {\cal K}$ s.t. $|r_0t^i_{r_0}|$+$|p_0t^j_{p_0}|$ is minimum, for $1 \le i,j,\le 2$\;
\eIf{no robot lies on $\overline{r_0d_1}$ (excluding point $d_1$ itself)}{
$Destination \leftarrow d_1$\;
}
{
$SafeRegion \leftarrow$ the part of the region inside the $SEC$ but outside $Cir(O, d'+\epsilon)$ that lies between $\overrightarrow{Or_0}$ and $\overrightarrow{Op_0}$ s.t. $\angle{r_0Op_0} \leq 180^o$\;
$d_0 \leftarrow$ a point $\in SafeRegion$, s.t. there exists no robots on $\overline{d_0d_1}$\;
$Destination \leftarrow d_0$\;
}
}
}
$r_i$ moves to $Destination$\;
\caption{MoveToDestination($r_i$)}
\label{Movetodestination}
\end{algorithm}

\paragraph{\bf Correctness of MoveToDestination($r_i$):}
The correctness of the algorithm is established by lemma \ref{lem-move-dest}.

\begin{lemma}
\label{lem-move-dest}
 Algorithm {\bf MoveToDestination($r_i$)} ensures collision-free movements of all robots to their final positions in finite time, without affecting the agreement on coordinate system.
\end{lemma}

\begin{figure}[!h]
\centering
\includegraphics{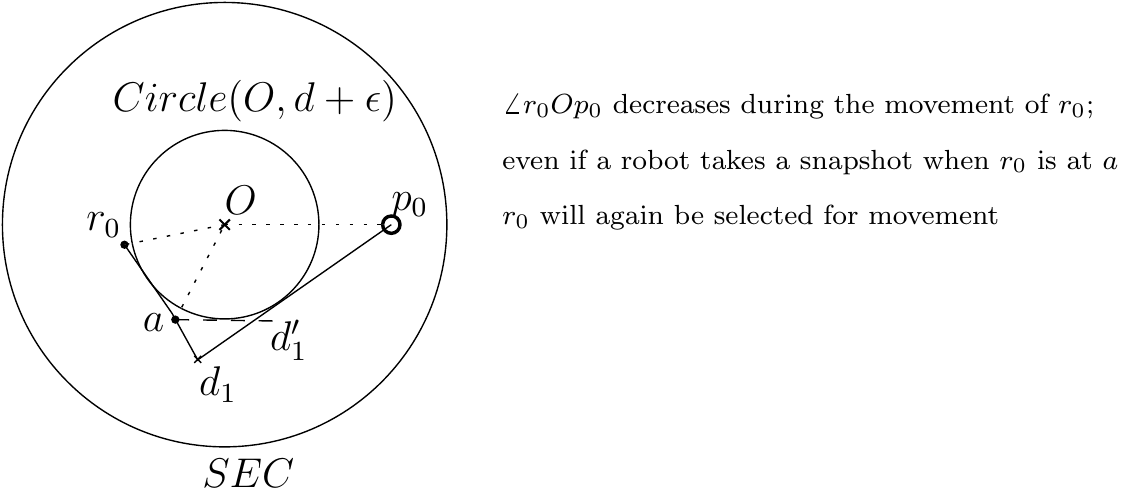}
\caption{An Example execution of {MoveToDestination($r_i$)}}
\label{MovToDest}
\end{figure}

\emph{Proof:}
 The movement of robots in $Free\bar{r}$ to final positions in $FreeP$, is designed in such a way that only one robot at a time moves (Figure \ref{MovToDest}). No robot in $Free\bar{r}$ gets closer to $O$ than $r_1$, which is already is at $p_1$. The algorithm selects a pair ($r_0 \in Free\bar{r}$, $p_0\in FreeP$) such that $\angle{r_0Op_0}$ is minimum. If the line $r_0p_0$ does not intersect $Cir(O, d'+\epsilon)$, then $r_0$ moves to $p_0$.
Else, if the line $r_0p_0$ intersects $Cir(O, d'+\epsilon)$, then $r_0$ travels through the tangents of $Cir(O, d'+\epsilon)$, while ensuring that the path to $p_0$ is the shortest.
This is achieved by computing a point $d_1$, s.t. $d_1$ is the intersection of tangents from $p_0$ and $r_0$ to $Cir(O, d'+\epsilon)$ and the distance from $r_0$ to $p_0$ via $d_1$ is minimum. $r_0$ then moves towards $d_1$.

Since, $r_0$'s last move was along $\overline{r_0d_1}$ towards of $d_1$, one of the following may happen:
(i) no robot takes snapshot till $r_0$ reaches $d_1$; (ii) other robots take snapshot during the movement of $r_0$ along $\overline{r_0d_1}$. In situation (ii), a robot which takes snapshot, will find $r_0$ at a point $a$ on $\overline{r_0d_1}$, other than $d_1$. $(aO, \angle{aOp_0})$ remains minimum during this movement. Hence, $r_0$ will again be selected for movement.  
 
The movement of $r_0$ does not result in any other robot getting closer to $p_0$ than $r_0$ or $r_0$ does not get closer to any point in ${\cal P}*$, other than $p_0$. Hence, in finite number of cycles, $r_0$ reaches $d_1$.
Moreover, when $r_0$ reaches $d_1$, it is selected for movement again. Thus, in every subsequent computation cycle the robot $r_0$ will be selected to move, until it reaches $p_0$. 
% The line segment $\overline{r_0p_0}$ is entirely within the $SafeRegion$. $r_0$ moves towards $p_0$.
% The robots in the region $SafeRegion$ already on their final positions do not move.
Hence, no collision occurs in the path of $r_0$, until $r_0$ reaches $p_0$
(either direct or via some $d_1$). $r_0$ reaches $p_0$ in finite number of cycles.

Throughout the execution of {\bf MoveToDestination($r_i$)}, no robot moves such that the $SEC$ changes. Hence, the agreement on $O$ i.e., the origin remains intact. The robots on $SEC$ and +ve $X$ axis is already is in ${\cal P}*$. Thus the axes and unit distance are also unchanged.
\qed

Now we present an algorithm, {\bf MoveOnBoundary($r_i$)}, for the movement strategy of the robots lying on the $SEC$. The algorithm assumes that the robots agree in coordinate system (using {\bf AgreementCoordinateSystem()}). 
In order to avoid collisions we define a configuration called {\it alternate configuration}.
\begin{defi}
 \label{defi-alt-conf}
 If there exists at least one pair $(r \in Free\bar{r}$, $p \in FreeP)$ on the $SEC$, such that  $p$ lies on the $SEC$ and $r$ can move to $p$ without passing through a filled final position, then the corresponding configuration is called an alternate configuration.
\end{defi}
In {\bf MoveOnBoundary($r_i$)} which is described next, we use three procedures.
\begin{itemize}
 \item $MoveOnCircle$($r_i$, $p'$) ensures that a robot reaches $p'$ moving strictly on $SEC$.
 
 \item $MoveEnsuringAlternate$($r_i$) is executed when only one robot on $SEC$ is at its final position on $SEC$. The function moves a robot in such a way that the alternate configuration is maintained. For example in Figure \ref{AltConfig}, $r_x$ moves to $B$ even though $A$ is nearer, just to ensure that the resulting configuration is alternate.
 
\begin{figure}[!h]
\centering
\includegraphics{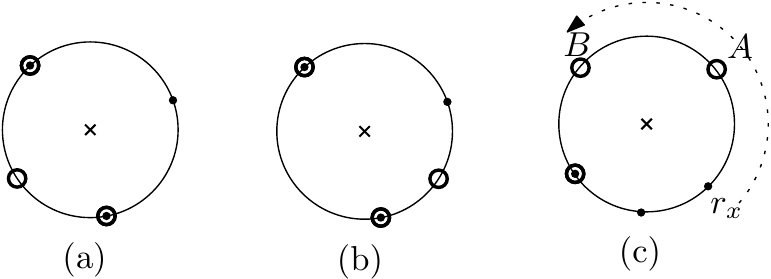}
\caption{(a)A non-alternate configuration, (b)An alternate configuration, (c)Example execution of \emph{moveEnsuringAlternate()}} 
\label{AltConfig}
\end{figure}

\item $GenerateAlternate$($r_i$) is executed when two robots on $SEC$ are at their final position on $SEC$ and the configuration is not alternate. The algorithm generates an alternate configuration. The function selects one robot $r_x \ne r_l$ ($r_l$: leader on $SEC$) out of the two robots in ${\cal P}_{Filled}$, s.t. $\angle{r_xOp'}$ ($\forall p' \in $(free points in ${\cal P}_{SEC}$)) is minimum and moves it along $SEC$ by a very small distance $\epsilon$ towards $p'$.
\end{itemize}

\begin{algorithm}[H]
\KwIn{$\bar{r}$: the set of robots, $\cal P$: the set of pattern points.}
\KwOut{ A configuration of robots where a robot on $SEC$ is in its final position.}

$\bar{r}_{SEC} \leftarrow$ \{$r_j$: $r_j$ is on the $SEC$\}\; 
${\cal P}_{SEC}  \leftarrow$ \{$p_k$: $p_k \in {\cal P}*$ and $p_k$ is on the $SEC$\}\;
${\cal P}_{Filled} \leftarrow$ \{$p_l$: $p_l \in {\cal P}*$, $p_l$ is on the $SEC$, and $p_l$  is occupied by a robot\}\;

\Switch{${\cal P}_{Filled}$}{

\Case{$|{\cal P}_{Filled}| = 0$:}{
Find a robot $r_n \in \bar{r}_{SEC}$ and a point $p' \in {\cal P}_{SEC}$ s.t. $|r_np'|$ is minimum\footnote{The ties are resolved using the lexicographic ordering.} \;
\eIf{$r_i  == r_n$}{
$MoveOnCircle$($r_i$, $p'$)\;}
{$r_i$ does not move\;}
}
\Case{$|{\cal P}_{Filled}| = 1$:}{
\eIf{$\exists$ $p' \in {\cal P}_{SEC}$ diametrically opposite to the point in ${\cal P}_{Filled}$}{
Find a robot $r_n$ s.t. distance $|r_np'|$ is minimum$^a$\;
\eIf{$r_i == r_n$}{
$MoveOnCircle$($r_i$, $p'$)\;}
{$r_i$ does not move\;}
}
{
$MoveEnsuringAlternate$($r_i$)\;
}
}
\Case{$|{\cal P}_{Filled}| = 2$:}{
\eIf{the two robots in ${\cal P}_{Filled}$ lie on a diameter of $SEC$}{
MoveToDestination($r_i$)\;
}
{
\eIf{configuration is alternate}{
Find a robot $r_n\in \bar{r}_{SEC}$ and point $p' \in {\cal P}_{SEC}$ s.t. $|r_np'|$ is minimum$^a$\;
\eIf{$r_i == r_n$}{
$MoveOnCircle$($r_i$, $p'$)\;}
{$r_i$ does not move\;}
}
{$GenerateAlternate$($r_i$)\;}
}
}

\Other{/* {$|{\cal P}_{Filled}| > 2$:}*/ \\
 MoveToDestination($r_i$)\;
}
}

\caption{MoveOnBoundary($r_i$)}
\normalsize
\label{Moveonboundary}
\end{algorithm}

\paragraph{\bf Correctness of the algorithm MoveOnBoundary($r_i$):}
The correctness of the algorithm is established by lemma \ref{lem-move-boundary}.
\begin{lemma}
\label{lem-move-boundary}
{\bf MoveOnBoundary($r_i$)} ensures that when a robot is moving on $SEC$, it does not collide with other robots and all points in ${\cal P}*$ on the $SEC$ are filled by robots in finite time. 
\end{lemma}
\emph{Proof:}
Algorithm {\bf MoveOnBoundary($r_i$)} maintains alternate configuration on the $SEC$, during its execution. It avoids collision between the robots. 
If the configuration is already alternate, then $MoveEnsuringAlternate(r_i)$ makes sure that the configuration will remain alternate. 
If the configuration is not alternate, then 
$GenerateAlternate(r_i)$ ensures the resulting configuration is an alternate configuration.
It also assures that only one robot on the $SEC$ occupies a final position on the $SEC$.
As a result, in the next cycle, the robots will execute Case $1$ of the algorithm. Since the configuration is already alternate configuration, the robot executes $MoveEnsuringAlternate(r_i)$, which will then result in 
(i) two robots are in their final positions on the $SEC$, and (ii) robots on boundary are in an alternate configuration. 
If the $SEC$ is in alternate configuration, in each cycle a robot on $SEC$ has a movement to its destination in ${\cal P}*$ on SEC. Since, the number of robots is finite and they complete their cycles in finite time, the points in ${\cal P}*$ on the $SEC$ will be filled by the robots in finite time.  
\qed

\vspace{30pt}
We have already presented all algorithms on which our pattern formation depends. Now we present the main algorithm, {\bf PatternFormation($r_i$)}.

\begin{algorithm}[H]
\small
\KwIn{$\bar{r}$: the set of robots, $\cal P$: the set of pattern points.}
\KwOut{ A configuration of robots where $r_i$ is at its final position.}
Agreement\_coordinate\_system()\;
% MoveOnBoundary($r_i$)\;
\Switch{$r_i$}{
\Case{$r_i = O$:}{
\eIf{$O \in {\cal P}*$}{
$r_i$ does not move\;
}
{$r_i$ moves to $p$ (a point on positive X-axis at a distance $\epsilon$ from $O$)\;
}
}
\Case{$r_i = r_1$:}{
\eIf{$( O \notin {\cal P}* )$}
{\eIf{$(\exists$ a robot at $O)$}{
$r_i$ does not move\;}
{$r_i$ moves to $O$\;}
}

{
\eIf{$(\exists$ a robot at $O)$}
{
 \eIf{$(r_i \neq p_1)$ and ( $\exists$ exactly one robot $r_i$ s.t. $0 < |Or_i| < (d+\epsilon)$)}
{$r_i$ moves to $p_1$\;}
{$r_i$ does not move\;}
}
{$r_i$ moves to $O$\;}

}
}
\Case{$r_i \in \bar{r}_{SEC}$:}{
\eIf{$((O \notin {\cal P}*)$ and $( \exists$ a robot at $O))$ or $((O \in {\cal P}*)$ and $( \nexists$ a robot at $O))$ or $(\exists$ more than one robot at a distance $ < (d+\epsilon))$ or $(r_{1} \neq p_1))$}{$r_i$ does not move\;
}
{
\eIf{ $(\exists p \in {\cal P}_{SEC}$ s.t. $\nexists$ a robot at $p)$ and $(\nexists$ a free robot strictly inside $SEC)$}{
MoveOnBoundary($r_i$)\;
}
{
 MoveToDestination($r_i$)\;
 }
 }
 }
 
\Other{
\eIf{$((O \notin {\cal P}*)$ and $(\exists$ a robot at $O))$ or $((O \in {\cal P}*)$ and $(\nexists$ a robot at $O))$}{
 $r_i$ does not move\;
}
{
\eIf{$\exists$ more than one robot at a distance $ < (d+\epsilon)$}{
 MoveRadiallyOut($r_i$)\;
}
{
\eIf{$r_{1} \neq p_1$}{
$r_i$ does not move\;
}
{
\eIf{$\exists$ $p \in {\cal P}_{SEC}$ s.t. $\nexists$ a robot at $p$}{
MoveToDestination($r_i$)\;
 }
 {
MoveToDestination($r_i$)\;
 }
 }
 }
 }
 }
 }
\caption{PatternFormation($r_i$)}
\normalsize
\label{Patternformation}
\end{algorithm}

\paragraph{\bf Correctness of the algorithm PatternFormation($r_i$):} 
The correctness of algorithm {\bf PatternFormation($r_i$)} is justified by following lemmas.

\begin{lemma}
\label{lem-origin-aree}
Once an agreement on origin ($O$) of the coordinate system is achieved, it remains invariant until the pattern is formed.
\end{lemma}

\emph{Proof:}
All robots agree on the center of the $SEC$, as their origin ($O$) of the coordinate system. No robot on the $SEC$ is allowed to make a movement such that the $SEC$ changes. Hence, $SEC$ remains intact throughout the execution of the algorithm. Therefore, $O$ also does not change throughout the execution of the algorithm. 
\qed

\begin{lemma}
\label{lem-axis-agree1}
 If $O$ $\notin$ ${\cal P}*$, the agreement on \emph{XY-axis} remains invariant until the pattern is formed.
\end{lemma}

\emph{Proof:}
The $X$ axis is decided by the ray $Or_{l}$ ($r_l$ is the leader on $SEC$ and in ${\cal P}*$). $r_l$ does not move.
Hence, the $X$ axis remains invariant.
$Y$ axis is decided by the $X$ axis and the final positions on boundary of $SEC$. Since $SEC$ remains invariant, the final positions on $SEC$ also remain invariant.
Hence, the $Y$ axis also remains invariant until the pattern is formed.
\qed
\begin{lemma}
\label{lem-axis-agree2}
 If $O$ $\in$ ${\cal P}*$, then the agreement on coordinate axes remains invariant from the point a robot reaches $O$ till the pattern is formed.
\end{lemma}

\emph{Proof:}
If $O$ $\in$ Final Positions, according to the algorithm $r_1$ moves to $O$. Following lemma \ref{lem-axis-agree1} the agreement on \emph{XY-axis} remains invariant until the pattern is formed.
\qed

Let us denote the initial configuration of the robots as $I_0$.
$I_1$ denotes configuration when $r_1$ occupies its final position at $p_1$.

\begin{lemma}
\label{lem-I0-I1}
 $I_0$ gets transformed to $I_1$ in finite number of cycles.
\end{lemma}

\emph{Proof:}
{\bf PatternFormation($r_i$)} ensures that $I_1$ is formed as follows: if $O \notin {\cal P}*$ and $\exists$ a robot $r_0$ at $O$, then $r_0$ moves to distance $\epsilon$ in the direction of positive $X$ axis. If $O \in {\cal P}*$ and $\nexists$ robot at $O$, a robot moves to origin. Subsequent execution of {\bf PatternFormation($r_i$)} finds a new $r_1$ which remains invariant because, every robot (other than $r_1$) inside or on a circle with radius $d$, moves to a distance $d+\epsilon$ from $O$. $r_1$ then occupies the position $p_1$.
Owing to the fact that the distance $d$ is finite, $I_1$ is formed in finite number of cycles. 
\qed

Consider another configuration $I_2$ defined as one in which the robot $r_1$ is at final position $p_1$ and all the final positions on the $SEC$ are occupied by some robots.

\begin{lemma}
\label{lem-I1-I2}
{\bf PatternFormation($r_i$)} transforms $I_1$ to $I_2$, in finite number of cycles.
\end{lemma}

\emph{Proof:}
According to algorithm {\bf PatternFormation($r_i$)} the robots in $\bar{r}_{SEC}$ do not move unless, either all position in ${\cal P}_{SEC}$ have been occupied or number of free robots strictly inside $SEC$ becomes zero.
This is assured because once $I_1$ is formed, the robots move in a specific order: First, the robots on $SEC$ fill the final positions in ${\cal P}_{SEC}$. Subsequently, if the number of free robots strictly inside $SEC$ becomes zero and the total number of free positions in ${\cal P}_{SEC} \neq 0$,
the free robots on $SEC$ follow algorithm {\bf MoveOnBoundary($r_i$)} to fill final positions on $SEC$, while keeping $SEC$ intact.
This continues until all the final positions on the $SEC$ are occupied by some robot. 
Hence, $I_2$ is formed form $I_1$ in finite number of cycles.
\qed

\begin{lemma}
\label{lem-I2-P}
{\bf PatternFormation($r_i$)} transforms $I_2$ to $\cal P$, in finite number of cycle.
\end{lemma}

\emph{Proof:}
Once $I_2$ is formed, free robots execute {\bf MoveToDestination($r_i$)} to reach the unoccupied final positions.
Robots already at their final position remain stationary. 

{\bf MoveToDestination($r_i$)} ensures robots reach their final positions in finite time. Hence, whatever unoccupied final positions exist in $I_2$ are filled by robots in a finite number of cycles and $\cal P$ is formed.
\qed

\begin{lemma}
\label{lem-sec-invariant}
 Even if the number of free robots strictly inside $SEC$ is less than the number of elements in ${\cal P}_{SEC}$, algorithm {\bf PatternFormation($r_i$)} executes successfully while keeping $SEC$ invariant.
\end{lemma}

\emph{Proof:}
If the number of free robots strictly inside $SEC$ is less than the number of elements in ${\cal P}_{SEC}$, {\bf PatternFormation($r_i$)} ensures that after configuration $I_0$ is formed, {\bf MoveToDestination($r_i$)} moves the free robots inside $SEC$ to the free points in ${\cal P}_{SEC}$. After which the algorithm {\bf MoveOnBoundary($r_i$)} is called and one of the following cases arises:

\begin{itemize}
\item There is no robot occupying a final position on the $SEC$:\\ 
In this case, only robot $r_n$ (the robot closest to a point $p'$ in ${\cal P}_{SEC}$) moves to $p'$. Note that, since nothing else has changed, in the next computation cycle a robot will execute Case $1$.

\item There is exactly one robot occupying a final position on the $SEC$:\\ 
In this case, first the algorithm checks if there exists a point in ${\cal P}_{SEC}$ which is diametrically opposite to the robot already occupying a final position. If this is true then the robot nearest to this diametrically opposite point moves to it. Subsequently the circle $SEC$ will remain intact even if the free robots on $SEC$ are allowed to move.
Otherwise, i.e., if no such diametrically opposite pattern point exists, the algorithm makes one robot move on the circle to a pattern point, while ensuring that the configuration remains alternate configuration.
Next, the robot will see that there are two robots occupying final positions on the $SEC$ and the next case will be executed.

\item There are exactly two robots occupying final positions on the $SEC$:\\
In this case, the robot first checks if two robots occupying the ${\cal P}*$ on $SEC$ form a diameter of $SEC$. If this is true, then the circle will remain invariant even if other free robots on the $SEC$ are allowed to move. Robots call algorithm {\bf MoveToDestination($r_i$)} (which ensures that the rest of the free robots occupy their final positions).
Otherwise, robot checks if the configuration is an alternate configuration. If it is true, the robot nearest to a pattern point on the boundary moves on the circle to towards it, else the robot call $GenerateAlternate(r_i)$, which moves one of the two robots (which is not on the $X$ axis) at final position (on the $SEC$) in such a way that an alternate configuration is formed. It also assures that the $X$ axis does not change. This alternate configuration has only one robot remaining at a final position on the $SEC$. In the next cycle the robot enters the previous case.

\item There are three or more robots occupying final positions on the $SEC$:\\
In this case {\bf MoveToDestination($r_i$)} can be called with rest of the free robots as argument. Here, we do not need to worry about keeping $SEC$ invariant (since there are already at least three stationary points on the boundary of $SEC$).
\end{itemize}

Under none of the above cases does the algorithm allow any changes to $SEC$. Hence $SEC$ remains invariant. Also since only the number of robots occupying final positions on $SEC$ change in a cycle, the next cycle (if there is one) also executes one of the above cases, and after some finite number of steps, all the robots on boundary occupy points from ${\cal P}_{SEC}$.
\qed

Finally we present theorem \ref{th:patternformation}.
\begin{theorem}
\label{th:patternformation}
 If a set of robots is in asymmetric configuration then an asymmetric pattern formation is possible by the robots even without agreement in common coordinate system.
\end{theorem}

\section{Conclusion}

In this paper we present an algorithm to form a given asymmetric pattern. This assumes no agreement on coordinate axes (SoD, orientation and scale). 
Although we assume the robots to be point, it is possible to extended the algorithm for pattern formation by transparent fat robots. The only thing we need to ensure in the case of fat robots is that at every step, the robots check if the minimum enclosing circle of their initial configuration is sufficiently large to ensure collision free movement. This can be done by making one robot on the boundary of the circle move away from the center of the circle until the diameter becomes large enough. The detailed work on pattern formation for fat robots may be an interesting continuation of this work. 

\bibliographystyle{plain}
\bibliography{Thesis}

\end{document}